\documentclass[sigconf]{acmart}

\usepackage{balance}

\usepackage[utf8]{inputenc}
\usepackage[T1]{fontenc}
\usepackage{graphicx}
\usepackage{textcomp}
\usepackage{footnote}
\usepackage{hyperref}
\usepackage{cleveref} 
\usepackage{longtable}
\usepackage{tabularx}
\usepackage{listings}
\usepackage{array}
\usepackage{amssymb}
\usepackage{subcaption}
\usepackage{multicol}

\newcolumntype{P}[1]{>{\centering\arraybackslash}p{#1}}
\lstdefinestyle{code_listing}{
  basicstyle=\footnotesize,
  breakatwhitespace=false,
  breaklines=true,
  captionpos=b,
  keepspaces=true,
  numbers=left,
  numbersep=5pt,
  showspaces=false,
  showstringspaces=false,
  showtabs=false,
  tabsize=2
}
\usepackage[english]{babel}

\newcommand{\NAME}{CORSICA}

\def\BibTeX{{\rm B\kern-.05em{\sc i\kern-.025em b}\kern-.08emT\kern-.1667em\lower.7ex\hbox{E}\kern-.125emX}}
    
\copyrightyear{2020} 
\acmYear{2020} 
\setcopyright{acmlicensed}
\acmConference[ASIA CCS '20]{Proceedings of the 15th ACM Asia Conference on Computer and Communications Security}{October 5--9, 2020}{Taipei, Taiwan}
\acmBooktitle{Proceedings of the 15th ACM Asia Conference on Computer and Communications Security (ASIA CCS '20), October 5--9, 2020, Taipei, Taiwan}
\acmPrice{15.00}
\acmDOI{10.1145/3320269.3372196}
\acmISBN{978-1-4503-6750-9/20/10}

\acmSubmissionID{asiac032}

\begin{document}

\fancyhead{}

\title{CORSICA: Cross-Origin Web Service Identification}

\author{Christian Dresen}
\email{c.dresen@fh-muenster.de}
\affiliation{\institution{Münster University of Applied Sciences}}

\author{Fabian Ising}
\email{f.ising@fh-muenster.de}
\affiliation{\institution{Münster University of Applied Sciences}}

\author{Damian Poddebniak}
\email{poddebniak@fh-muenster.de}
\affiliation{\institution{Münster University of Applied Sciences}}

\author{Tobias Kappert}
\email{tobias.kappert@fh-muenster.de}
\affiliation{\institution{Münster University of Applied Sciences}}

\author{Thorsten Holz}
\email{thorsten.holz@rub.de}
\affiliation{\institution{Ruhr-University Bochum}}

\author{Sebastian Schinzel}
\email{schinzel@fh-muenster.de}
\affiliation{\institution{Münster University of Applied Sciences}}

\renewcommand{\shortauthors}{Dresen and Ising, et al.}

%

%
\begin{abstract}
Vulnerabilities in private networks are difficult to detect for attackers outside of the network. While there are known methods for port scanning internal hosts that work by luring unwitting internal users to an external web page that hosts malicious JavaScript code, no such method for detailed and precise service identification is known. The reason is that the Same Origin Policy (SOP) prevents access to HTTP responses of other origins by default.

We perform a structured analysis of loopholes in the SOP that can be used to identify web applications across network boundaries. For this, we analyze HTML5, CSS, and JavaScript features of standard-compliant web browsers that may leak sensitive information about cross-origin content. The results reveal several novel techniques, including leaking JavaScript function names or styles of cross-origin requests that are available in all common browsers. 

We implement and test these techniques in a tool called \NAME{}. It can successfully identify 31 of 42 (74\%) of web services running on different IoT devices as well as the version numbers of the four most widely used content management systems WordPress, Drupal, Joomla, and TYPO3. \NAME{} can also determine the patch level on average down to three versions (WordPress), six versions (Drupal), two versions (Joomla), and four versions (TYPO3) with only ten requests on average. Furthermore, \NAME{} is able to identify 48 WordPress plugins containing 65 vulnerabilities.

Finally, we analyze mitigation strategies and show that the proposed but not yet implemented strategies \emph{Cross-Origin Resource Policy (CORP)} and \emph{Sec-Metadata} would prevent our identification techniques.
\end{abstract}

%
%

\ccsdesc[500]{Security and privacy~Web application security}
\ccsdesc[300]{Security and privacy~Network security}
\ccsdesc[100]{Security and privacy~Browser security}

\ccsdesc[500]{Computer systems organization~Embedded systems}
\ccsdesc[300]{Computer systems organization~Redundancy}
\ccsdesc{Computer systems organization~Robotics}
\ccsdesc[100]{Networks~Network reliability}

%
\keywords{SOP; JavaScript; Service Identification; Fingerprinting; Web Security; Perimeter Security}

%

%
\maketitle

\setcounter{page}{1}

\section{Introduction}

Network segregation remains a cornerstone of Internet security as it often separates insecure services within private networks (\emph{Intranet}) from the public Internet. For example, a firewall acts as a network perimeter that filters packets between both networks, often in a way that outgoing connections are allowed and incoming connections blocked. Setups with network address translation (\emph{NAT}) commonly show this behavior. Administrators focus on securing Internet-facing services but often treat Intranet applications with minor importance. In consequence, Intranet applications are often not patched with the latest security updates or use insecure configuration settings. This leads to situations where Intranet applications may contain well-known vulnerabilities that are easy to exploit using common attack frameworks \cite{metasploit, beef}. Exploits against web applications play a unique role here because attacks like Cross-Site Request Forgery (CSRF), Cross-Site Scripting, or SQL Injection can be used to cross perimeters. For example, a web attacker can create an \textit{Internet} website containing exploits and lure unsuspecting users within the private network to visit that page. The victim's browser then attacks a target within the private network.

\begin{figure*}[!ht]
\centering
    \includegraphics[width=0.85\textwidth]{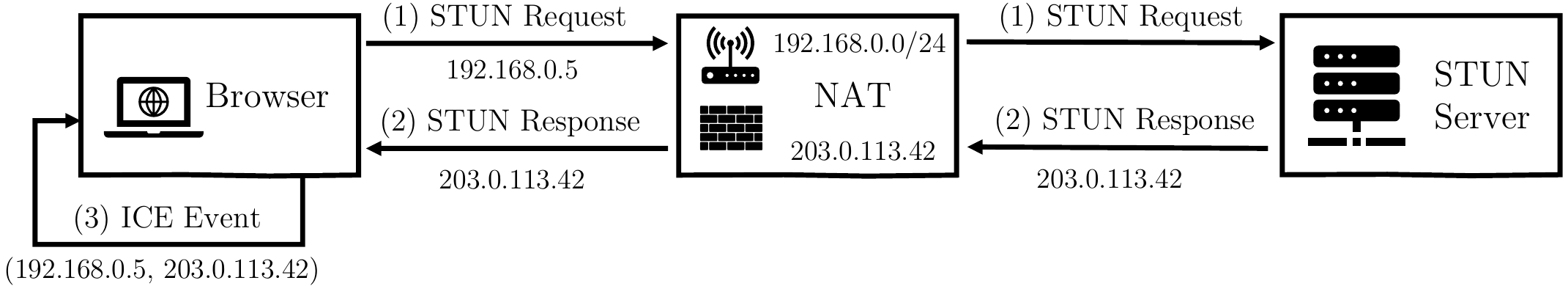}
    \caption{Simplified STUN process: A client behind a NAT sends a STUN request to a STUN server and receives its Internet-facing IP address in response.}
    \label{fig:stun}
\end{figure*}

From a web attacker's viewpoint, she has the problem of learning detailed information about the victim's network in order to perform such an attack successfully. \emph{At which IP/port within the private network runs a vulnerable web application with known exploits?}

There are known attacks that randomly try known exploits against IP addresses in private networks \cite{jakoblell-dns-csrf, kafeine-dns-csrf}, which is effective for compromising many random hosts but is not suitable for targeted attacks. While it is known to be possible to perform port scans in a cross-origin fashion using JavaScript~\cite{pingjs}, there is no such method to learn detailed information about web applications like type and version. Existing web service identification tools do not work from browsers and across origins. This information shortage about the victim's network makes targeted attacks against web applications in private networks difficult.

This leads to our main research question:
\emph{Is it possible to perform precise web service identification across network perimeters that use standard browser functionality and that work across origins?}

We perform a structured analysis of standard-conforming CSS, HTML5, and JavaScript functionality and introduce several techniques that can be used to create unique identifiers of web applications and that work across origins. These identifiers are used to query a database of known identifiers to learn the type and version of web applications.

We implemented the proposed techniques in a tool called \NAME{}\footnote{https://github.com/FHMS-ITS/CORSICA} and measured its effectiveness against a testbed of widely used web services and different IoT services. The analysis shows that \NAME{} can successfully detect the type and version of web applications running on $250$ IoT devices, $122$ different versions of content management systems, and $48$ WordPress plugins with known, remotely exploitable vulnerabilities.

\noindent In summary, we make the following contributions in this paper:
\begin{itemize}
    \item We present novel techniques that allow identifying web-based services using standard browser features and that work cross-origin (\Cref{sec:identify}).
    
    \item We describe a method to automatically learn feature vectors for identifying a web-based service (\Cref{sec:learn-service-feature-vectors}).
    
    \item We implement the techniques in a tool called \NAME{} and show that it can successfully identify 154 out of 250 tested IoT systems running web applications. It can identify patch level versions of content management systems on average down to three versions (WordPress), six versions (Drupal), two versions (Joomla), and four versions (TYPO3) as well as 48 out of 600 WordPress plugins containing known remotely exploitable vulnerabilities (\Cref{sec:evaluation}).

    \item Finally, we discuss possible strategies that prevent the presented identification techniques (\Cref{sec:countermeasures}).
    
\end{itemize}

\section{Background and Related Work}

In its current state, the Internet is not a single large network of devices, it rather is a collection of multiple smaller networks and devices. Smaller networks are often encapsulated and cannot be directly reached from the outside. However, collaboration in the form of resource sharing between services on different devices and networks is a desired feature of the Internet. While these features are commonly used by many web services for collaboration, they allow a non-negligible attack surface if implemented insecurely.

\subsection{Perimeter Security}

Today, services on private and corporate networks are usually hidden behind a perimeter in the form
of firewalls or Network Address Translators (NATs). While firewalls actively protect services from
external access by blocking specific ports and traffic, NATs usually provides the same feature
passively.

NAT was standardized in the 1990s by the IETF \cite{RFC1631} in response to concerns about the
exhaustion of the available IPv4 address space. While the Internet was initially designed with the
assumption that every device with access to the Internet has its own IP address, NAT enables the
reuse of specific address blocks for internal networks. NAT boxes enable this reuse by translating internal
IP addresses to a single Internet-facing IP address, effectively blocking direct access to devices
on the internal networks from the Internet.

Nevertheless, it is possible to learn certain information about the network using JavaScript inside a browser in a NAT environment. For example, a malicious website can perform network scans revealing available hosts as well as, to a certain extent, open ports. An example of this is \textit{pingjs} by Jonathan Frederic \cite{pingjs} that uses \texttt{onload} and \texttt{onerror} events inside JavaScript to discover hosts inside the network.
Furthermore, as NAT breaks desired functionality of the Internet (e.g., peer-to-peer connections between clients
on different networks), several workarounds, called NAT traversal strategies, have been proposed.
One of these proposals are the \emph{Session Traversal Utilities for NAT} (STUN, see Figure~\ref{fig:stun}). STUN requires a STUN
server outside of the internal network of a client. First, the client sends a STUN request to the
server.  As the client is located behind NAT, its internal IP address is translated to the Internet-facing
IP address before it reaches the STUN server. The STUN server then returns a STUN response
containing the contacting public IP address to the client \cite{RFC5389}. Additionally,
if STUN is used from JavaScript, the browser will provide the internal IP address to the JavaScript context.

\subsection{Same-Origin Policy in Browsers}
\label{sec:background-sop}
The {\em Same-Origin Policy (SOP)} restricts the interactions of scripts between different origins. This means that the resource access between these origins is restricted~\cite{mozilla-same-origin}. Examples for resources are images,
scripts, styles, or frames.  Netscape introduced the SOP in 1995 alongside JavaScript and the
Document Object Model (DOM) \cite{tangled-web}. Interaction of web resources from multiple origins
is also called \emph{cross-origin access}. In general, the SOP allows to request and embed resources from
other origins. However, it forbids read access to, and interaction with the content received from
another origin.  Even though the SOP has been a staple of web security for many
years, to this day, it is not formally defined and not implemented consistently in different browsers
\cite{schwenksop2018}.

As described above, embedding resources like images, style sheets, and even whole pages into a website is typically allowed. 
For websites, it is necessary to access some information about the embedded cross-origin resource to place them correctly inside their DOM. Therefore some information must be made available.
For example, a script running in the origin \textit{attacker.org} can
request a picture from the origin \textit{target.org}. Therefore \textit{attacker.org} can embed it into her own DOM and can read the dimensions of the image to format the website accordingly. However, it cannot interact with the actual contents of the response, including access to any pixel data of the image \cite{mozilla-same-origin}.

\subsection{Cross-Site Request Forgery}
\begin{figure*}[t!]
    \centering
    \includegraphics[width=\textwidth]{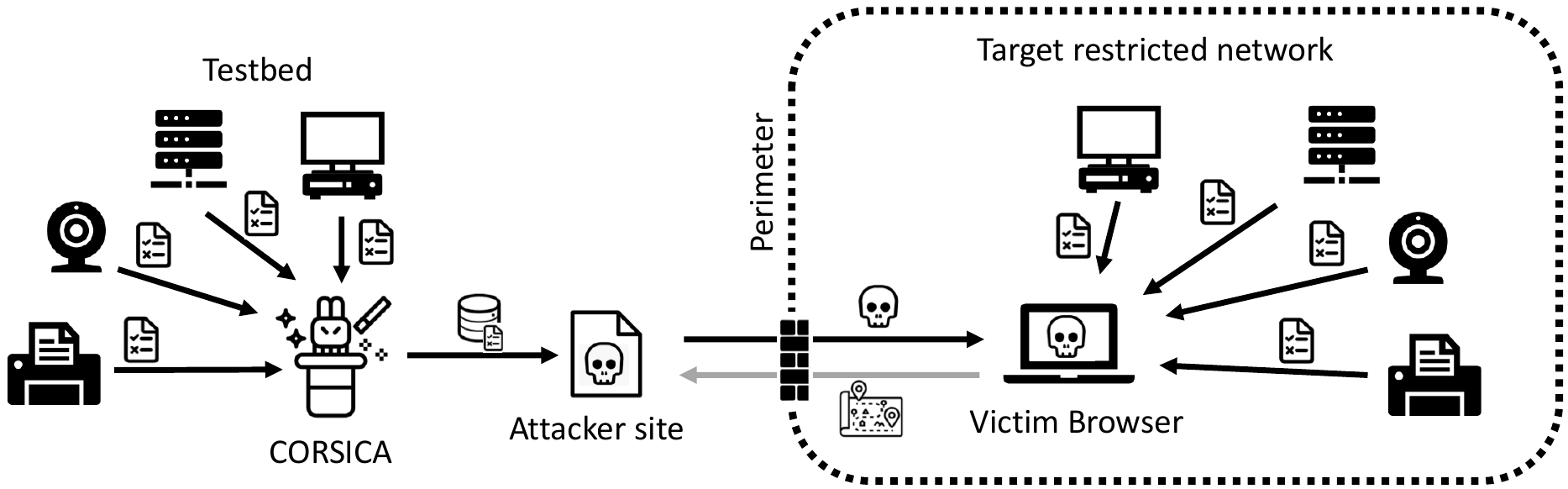}

    \caption{The architecture of the cross-origin web service identification tool \NAME{}. }
    \label{fig:widen}
\end{figure*}

Cross-Site Request Forgery (CSRF) is a type of attack against web applications that exists since the
1990s but got its name in 2001 from Peter Watkins. CSRF attacks trick a user into performing
actions of an attacker's choosing by leveraging the victim's browser \cite{burns2007}.
For example, an unassuming user's browser that is lured onto a website controlled by an attacker
could be tricked into making a request to their webmail provider trying to install a filter to
forward all e-mails to the attacker. Assuming the victim is logged into their webmail account, the
victim's browser sends the session cookie with the attacker's request, and the requested filter is
applied. This exact attack was performed against Gmail in 2007 \cite{gnucitizen2007}.

CSRF attacks are possible because modern web applications or websites are often required to be able to send HTTP requests to any network address. This ability, however, allows an attacker that controls a website visited
by the victim's browser to use resources not otherwise under her control. These resources include
leveraging the browser state (e.g., by using authentication cookies to perform privileged actions
under the victim's identity) and network connectivity (e.g., accessing network resources behind
a perimeter such as a firewall) \cite{barth2008}.
A common defense mechanism against CSRF attacks are so-called CSRF tokens that are bound to a victim's
session and need to be sent with every request made. Only if the token can be correctly verified,
the requested action is performed \cite{barth2008}.

\subsection{Related Work}
Substantial work on identifying Internet-reachable services was done in the past, and in the following, we survey work related to our approach. Konzina et al. collected characteristic information on web
applications in 2009 by identifying linked applications-internal URLs and forms \cite{Kozina2009}. However, their tool requires the attacker to be on the same network to conduct a scan.

Acar et al. analyzed web-based attacks against local IoT devices behind NATs using a malicious website \cite{Acar2018}. Their work mainly relies on circumventing the same-origin policy by exploiting HTML \texttt{MediaErrors} and DNS rebinding in web browsers. Their attack will not work with most modern browsers, because they do not generate \texttt{MediaErrors} anymore \cite{bugreport-mediaerror-chrome,bugreport-mediaerror-firefox}. Furthermore, DNS Rebinding to local IP addresses is blocked by many modern home routers \cite{fritzbox-dns-rebind-protection, netgate-dns-rebind-protection}, making this technique unreliable. Compared to this, \NAME{} stays within the borders of the SOP and thus does not rely on DNS rebinding.

Stamm et al. presented attacks using a malicious web page to detect and manipulate home routers \cite{Stamm2017}. The authors build upon this to allow access to the internal network by reconfiguring the victim's DNS server and using these attack vectors to persist. The limitations presented in their work, however,
do not allow for generic fingerprinting of Intranets.

Additionally, much work was done on fingerprinting JavaScript files. In 2012, Blanc et al. described a procedure to characterize obfuscated
JavaScript using abstract syntax trees \cite{Blanc2012}. Even though this would also help fingerprint non-obfuscated JavaScript files,
such a technique would require unrestricted access to the source code, which is prevented by the SOP.
In 2008, Johns documented state-of-the-art techniques in Java\-Script malware \cite{Johns2008}. More specifically, he presented readable attributes for
cross-origin resources that are dynamically loaded on a website. {\NAME} builds upon and extends these loopholes in the SOP to build a web-based fingerprinter and scanner for networks across perimeters.

In 2015, Frederic published a JavaScript-based ping script \textit{pingjs} on GitHub \cite{pingjs}. To determine if a host on the network is online, an \texttt{img} element is created, and the resource URL is set to the IP address of the host. The \texttt{onload} and \texttt{onerror} events \cite{w3c-onload, w3c-onerror} of the image element are used in conjunction with a timeout to determine if the host is up. In March 2019, Bergbom published a research report regarding the described attack scenario. It uses similar techniques but is limited to one media type: images \cite{forcepoint-research-report}.

Subtle differences in HTML pages were found to not only leak information about type and version of the web application but may also correlate with confidential information that is leaked through a storage-based side channel \cite{10.1007/978-3-642-21424-04}.

It is well-known that using JavaScript enables a website or web service to fingerprint the user's web browser using several techniques \cite{Upathilake2015}. As these fingerprinting techniques target the web browser of the user, it raises privacy concerns about enabling user tracking. These techniques rely -- among other aspects -- on the fact that every browser interprets styles and scripts differently. In the case of this work, this browser-specific behavior is rather obstructive and complementary to our goals.

\section{Attacker Model and Overview}
\label{sec:attacker-model}

The goal of the attacker we consider in this paper is to attack web services that are hidden behind security perimeters, e.g., on private networks behind firewalls or NAT devices. To gain access to this network, the attacker lures unwitting users located within the private network to visit an attacker-controlled Internet website. We assume that the victim uses a standards-compliant web browser that executes the JavaScript code from the attacker's website. The browser strictly adheres to the Same-Origin Policy when executing the JavaScript code.

For service identification, the attacker needs to scan the private network and identify the type and possibly version of web services. This information is key to find known security vulnerabilities that can be exploited to conduct the actual attack. The winning condition is reached when the attacker learns detailed information about the type, version, port, and IP address of a web service that has known security vulnerabilities but is hidden behind a security perimeter. Note that the actual attack against this web service after identification is out of scope for this work.

\autoref{fig:widen} shows the overall process we follow in this paper. At first, \NAME{} queries a testbed of web applications with known security vulnerabilities to generate feature vectors. Those feature vectors are built into a classification engine that is published in a JavaScript-based attacker website on a public webserver. When a client inside the network visits the page, \NAME{} is executed in the client's browser and discovers and identifies existing devices and applications in the restricted network. \NAME{} then exports a map of discovered and identified services within the restricted network.

\section{Identifying Services across Perimeters}
\label{sec:identify}

\begin{figure*}[t!]
    \centering
    \includegraphics{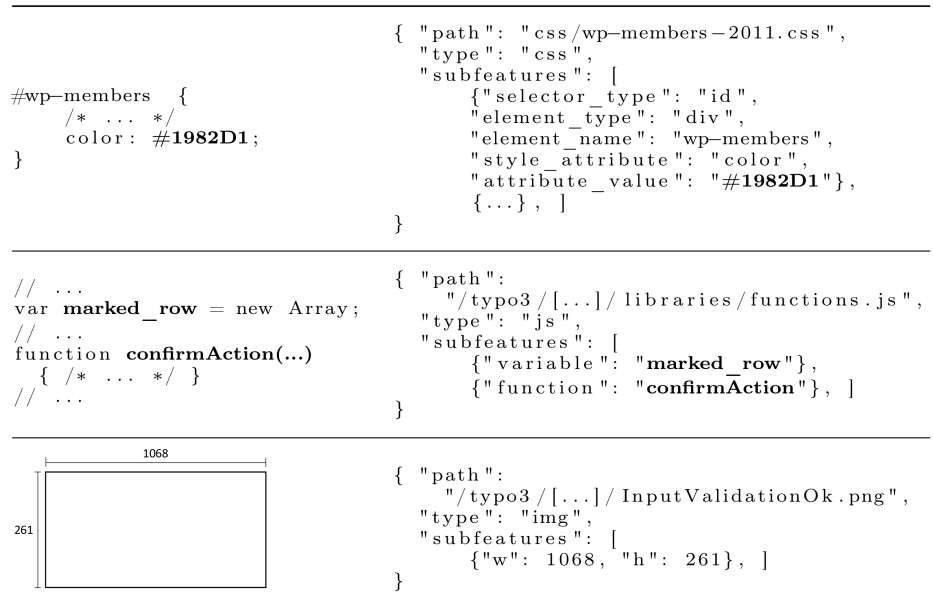}
    \caption{Features representing a style sheet, a JavaScript and an image file from the CMS: Wordpress and TYPO3}
    \label{fig:feature-ex}
\end{figure*}

Websites often include code and resources into their own DOM, either from a content delivery network (CDN) or from other domains.
As described in Section~\ref{sec:background-sop}, this is explicitly allowed by the SOP, as it prevents Cross-Origin reads but allows the inclusion of code (i.e., JavaScript or StyleSheets) and resources (i.e., images or videos).

Although the browser tries to reveal as little information as possible about a loaded resource, some information is ultimately necessary. For example, a web page needs to know the width and height of an included image to format a web page correctly. In other words, the SOP tries to be just permissive enough not to break benign features of the web.
However, these pieces of information combined allow the precise identification of services.

\subsection{Loopholes in the SOP}
\label{sec:loopholes}

When a website loads resources, it can define multiple JavaScript handlers,
particularly \texttt{onload} and \texttt{onerror}. Those handlers are designed to serve information
to JavaScript concerning the loading state and if a loading error has occurred. In most cases, this
information is not passed to JavaScript for cross-origin requests. However, this behavior is inconsistent for different browsers and resource types.
While these events are essential to render websites embedding external image files, they are usually not relayed for arbitrary media files like PDF files and websites loaded in frames. This is to prevent a website from checking the existence of these files across origins.
As shown in \autoref{tbl-filetypes}, the vast majority of accessible files within web services are JavaScript, Style Sheet, or image files.

\begin{table}[ht]
\begin{center}
\caption{File types within web services inside our database.}
\label{tbl-filetypes}
\begin{tabularx}{\linewidth}{p{1.2cm}>{\raggedleft\arraybackslash}X}
    \toprule
    \textbf{Filetype} & \textbf{Number of files} \\ 
    \midrule
    .php:  & 1792301 \\
.gif: & 657480 \\
.png: & 434389 \\
.js: & 380695 \\
.css: & 152453 \\
.yml: & 146461 \\
.html: & 144020 \\
.rst: & 135629 \\
.xml:&  81685 \\
.svg: &  71222 \\

    \bottomrule
\end{tabularx}
\end{center}
\end{table}

\paragraph{Images.} 
Although pixel data is generally not available\footnote{This also includes functions on the image, e.g. filters, as they are known to reveal information about the image data \cite{context-pixel-perfect-timing,evonide-css3-features}.}, the image dimensions are crucial for correct formatting. Thus, the width and height of an image are always available across origins. This information allows checking the presence of an image as well as distinguishing between two images with identical file paths but different dimensions.

\paragraph{Cascading Stylesheets.}
Even though the content of a style sheet is not directly available to the embedding document, it is impossible to hide the formatting from the embedding web page.
Thus, after applying a style to an object in the DOM, the resulting attributes can be read. As a result, the embedding website can reconstruct a loaded style sheet to a considerable extent.

The example style sheet directive displayed in \autoref{fig:feature-ex} defines directives for an element with the id \texttt{wp-members}. Information on the directive can be accessed to some extent by including the style sheet into an HTML page and creating an element with the id \texttt{wp-members}. After creation, the browser will automatically apply the style values to the elements under the condition that the style sheet was loaded correctly. In this specific case, we would expect the browser to set the \texttt{color} attribute to \texttt{\#1982D1}.
Fortunately, we are allowed to read those style values from any element we have created with JavaScript using the function \texttt{getComputedStyle}. This function returns the value of a specific style attribute the browser has applied to an element. 
The applied values can be compared with the expected results to determine if a style sheet file is present on the device and contains the expected directives.

\paragraph{JavaScript.}
Remote JavaScript is necessarily accessible as if it was loaded from the same origin (with some restrictions regarding the visibility of syntax errors).
Although variables and functions cannot be listed via JavaScript, their existence can be checked. Furthermore, the values of defined variables can be read, and the embedding script can call functions. Lekies et al. \cite{190992} described attacks based on this.

Including a \texttt{<script>}-tag into an HTML page with the source attribute directing to the JavaScript file to be checked makes all functions and global variables accessible from a global context. Additionally, the output of functions or operations inside a JavaScript file is precalculated where possible.
As a result, the presence of functions and variables as well as their value can be checked using JavaScript.
Interestingly, calling \texttt{.toString()} on a function even returns the source code, including comments \cite{mozilla-function-tostring}.

\paragraph{Other Resources.}
Websites can also embed other types of media files, for example, audio and video files. For videos, the dimensions (\texttt{.videoHeight} and
\texttt{.videoWidth}) and duration is available (\texttt{.duration}).

\subsection{Web-based Service Identification}

\label{sec:web-based-service-identification}

Each service is identified by a set of files (i.e., features). A single feature consists of one or more subfeatures. Subfeatures represent actual attributes that can be extracted from a file and are necessary to identify the file using the techniques described in Section~\ref{sec:loopholes}.
\autoref{fig:feature-ex} shows example features representing a CSS, a JavaScript, and an image file.

\begin{figure*}[t!]
\centering
\begin{minipage}[b]{.9\columnwidth}
    \includegraphics[width=\columnwidth]{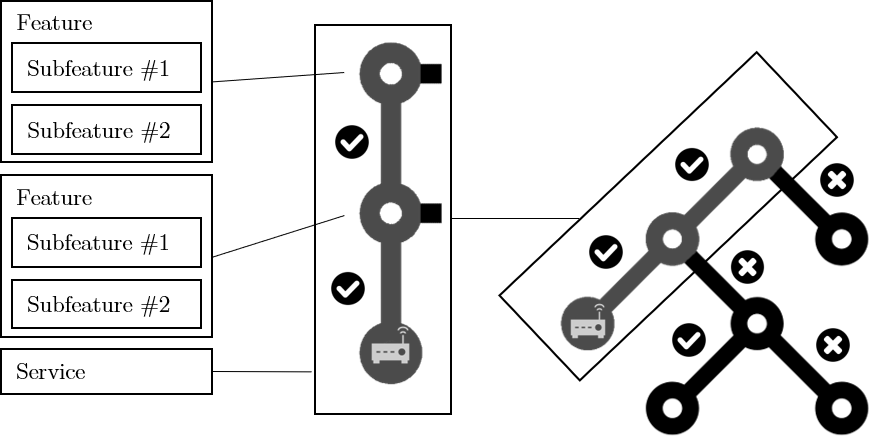}
    \caption{Example identification tree. A single path in the tree identifies a service uniquely. Each node represents a
    feature (file), for which the value can be tested by checking the subfeatures.}
    \label{fig:tree-features}
\end{minipage}
\qquad\qquad
\begin{minipage}[b]{.9\columnwidth}
\includegraphics[width=\columnwidth]{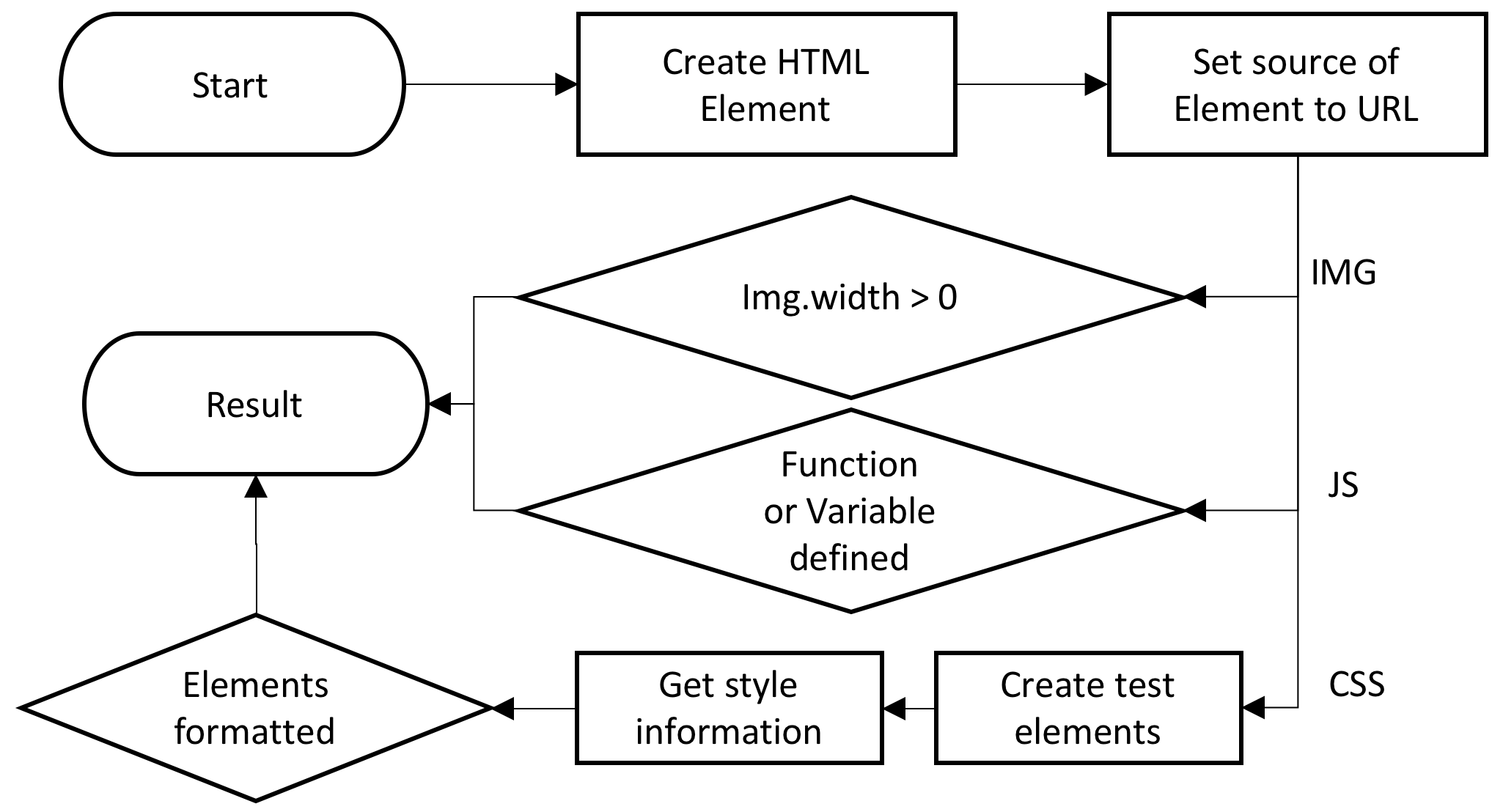}
\caption{Diagram of the identification process for features for each file type.}
\label{fig:feature-identification-process}
\end{minipage}
\end{figure*}

Once the feature vectors are generated, they can be used to identify a service using JavaScript. To
achieve this, a process based on the decision tree shown in \autoref{fig:tree-features} is used. At every node of the tree, a single feature 
of a service group is checked, and the next node is chosen according to the result. This will be repeated until a leaf node is reached, 
and the service is identified as uniquely as possible.

The process to check a single feature is described in \autoref{fig:feature-identification-process}.
To check a single feature, an HTML-Element corresponding to the file type is created on the test page: \texttt{<script>} for JavaScript, \texttt{<link>} for CSS, and \texttt{<img>} for images. In the next step, the \texttt{src} attribute of the created element is set to the URL of the file to identify.
When it is loaded, the techniques described in Section~\ref{sec:loopholes} are used to identify the file. For images, the image size is checked and, for JavaScript, all expected variables and functions are checked for their existence. For style sheets, there is an additional step to be done. Here we need to add test elements to our test page that can be formatted to check the directives. Those test elements must have the right class or id and the correct type. Once a test element is created, we can observe if the test element is formatted as expected. This can be done by using the JavaScript function \texttt{getComputedStyle} or the attribute \texttt{currentStyle}, depending on the browser used. After checking a directive, the element is removed, and a new test element is created to prevent
the page from slowing down due to excessive amounts of elements. After all subfeatures of a feature are tested, the created element from the beginning is removed from the test page.

\section{Learning Service Feature Vectors}
\label{sec:learn-service-feature-vectors}

Before identifying web-based services across perimeters, feature vectors need to be
collected. This step is necessary because the attacker needs to know which features to
collect and which vectors to compare the collected feature vectors to. In this section, we describe the
offline setup phase of the identification process.

\subsection{Creating a Service Corpus}
\label{sec:creating-a-service-corpus}
Files containing usable attributes for service identification can be retrieved using three different approaches depending on the availability of the needed data: (1) using the installation files of a service, (2) using a firmware image of a device, or (3) crawling an online installation.
As not all approaches are possible for every service, we made sure that the output of the three approaches is compatible, and that the results can be combined.

\paragraph{Installation files of Services.}
Obtaining installation files of a service is often possible from the vendor's or distributor's web page. After downloading and unpacking, we use the obtained files for the feature vector generation.

\paragraph{Firmware Images of Devices.}
Firmware images of IoT devices often contain Linux based file systems and can be unpacked to access the files. If the  downloading and unpacking part is successful, we can find the root directory of the web server and extract all relevant files, while preserving the directory paths as well as the real web path relative to the webroot on the device. This approach nevertheless has a couple of caveats. First, not all firmware images can be successfully unpacked. Second, many vendors do not provide public access to their firmware images but implement an automated update process on the individual embedded devices. This complicates access to those images.

\paragraph{Crawl Online Installations.}
The third approach uses a crawler to obtain the relevant files from live systems.
We initialize the crawler with the IP address and the port of a running web service or the web interface of a target device. Starting at the index page of the web service or interface, the crawler tries to find all relevant files accessible from the crawler's position. Those can be, for example, linked in the HTML files as well as inside JavaScript and style sheet files served by the server.
The downside of this approach is that the number of files the crawler can access is limited to files linked in publicly
accessible documents. In some cases, relevant files are publicly accessible, but only pages inside protected areas link to them, which the crawler cannot reach. This results in weaker feature vectors in comparison to those generated by the first two approaches.

\subsection{Extracting Feature Vectors}
\label{sec:extract_feature_vectors}

After creating the service corpus, we can begin to generate feature vectors, as described in \autoref{sec:web-based-service-identification}.

An image file is represented by a feature divided into two subfeatures -- the attributes width and height. This information can be extracted using image processing libraries.

JavaScript and CSS files result in more than two possible subfeatures as such files offer
more attributes we can access from JavaScript. To extract the subfeatures from
those files, we use a parser.
While extracting subfeatures for style sheets, we have to consider the different types of possible style definitions, such as defining attributes for all elements of a specific type and elements with IDs or classes. An example is described in \autoref{sec:web-based-service-identification}.

The subfeatures for JavaScript are functions or variables which are declared and accessible from a global context.
Besides a simple test of an existing function or variable, we can call the \texttt{toString} function of JavaScript on
any defined function with a name to get the source code, including comments, despite the SOP \cite{mozilla-function-tostring}. The basis for that can be extracted using the Selenium framework.

The resulting file identification features are combined to feature vectors. Those feature vectors are used to identify full services or single service components.

\subsection{Tree Construction and Optimization}
\label{sec:optimization}

The construction of feature vectors is only the first step in being able to identify services efficiently. While every
identified IP address/port pair could be checked against the whole service corpus, this process would be slow and error-prone
due to varying browser behavior. Therefore, first, the feature vectors will be normalized for behavior in different browsers, and after that, a tree will be constructed from the normalized corpus to optimize the identification process.

\begin{figure*}[t]
    \centering
    \includegraphics[width=0.9\textwidth]{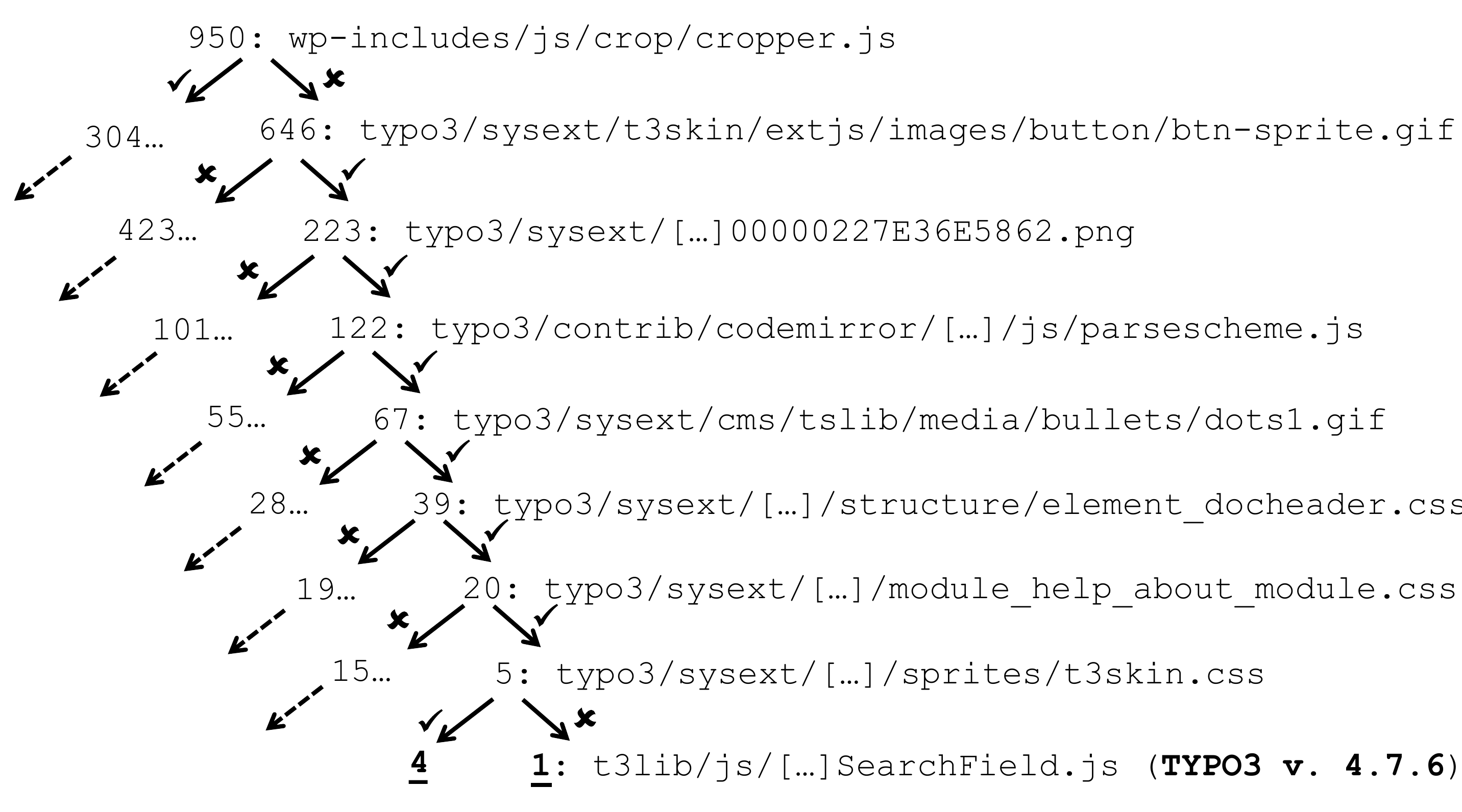}

    \caption{A path in the decision tree that leads to TYPO3 version 4.7.6. The numbers denote the amount of possible versions left and the path is the resource to be requested to move forward in the decision tree.}
    \label{fig:decision-tree}
\end{figure*}

\paragraph{Normalizing the Feature Vectors}
The identification of images works the same way in every browser. Unfortunately, this does not apply to JavaScript and style sheet files. The process of file identification shown in \autoref{fig:feature-identification-process} relies on the browser executing JavaScript or applying style directives and comparing the result with expected values. However, different browsers will interpret style sheets as well as JavaScript code differently, and, therefore, some of the generated feature vectors will not work correctly. Additionally, JavaScript files sometimes require other JavaScript code to be executed before to work, i.e., if functions rely on frameworks like jQuery. To deal with this, we have implemented an automated test suite based on the Selenium framework \cite{seleniumhq} supporting Chrome and Firefox. This test suite uses both browsers to test all available identification vector from the database against the files in question and marks all subfeatures that could not be recognized correctly, constructing normalized feature vectors that can be recognized in both browsers. Furthermore, this enables us to build separate feature vectors for each supported browser.

\paragraph{Tree construction}
The speed of the identification is crucial, as it can only run as long as the attacker's website is open in the victim's browser. As a browser only allows a limited number of parallel requests -- mostly between six and ten -- the time needed to identify a device or application is directly related to the number of required requests. To optimize that number, we use a decision tree-based approach when identifying services. 

To build an optimized tree, we identify service groups from the corpus that share features, e.g., devices by the same manufacturer that share a company logo. A service can be part of multiple service groups as it can share common features with multiple services that do not necessarily share the same features. Subsequently, we construct the tree by creating nodes to distinguish between services designed to split the remaining services in half with each level of the tree. Furthermore, the nodes are sorted by the frequency of services to increase the probability of traversing the tree faster. The resulting tree contains all services in the corpus. However, not all can be identified uniquely -- i.e., are represented by a leaf node in the tree -- because some service groups are indistinguishable by the extracted features.

\autoref{fig:decision-tree} shows an example of the path used to identify TYPO3 version 4.7.6. Each node represents a feature and therefore, a file requested by our test page. The first requested path is \path{wp-includes/js/crop/cropper.js}, which does not exist for TYPO3 v. 4.7.6. The second path is \texttt{btn-sprite.gif}, which exists. The numbers in each node denote the number of possible device versions left. The last requested path is \texttt{SearchField.js}, which exists and thus identifies TYPO3 version 4.7.6.

\paragraph{Optimizing the Identification Process.}
We distinguish between two identification processes:
The full service identification and  the component identification. The full service identification has exactly one result, while the component identification can lead to multiple independent results as an application can consist of multiple components (i.e., plugins). For the full service identification, the number of requests necessary to identify a service is critical.

\section{Evaluation}
\label{sec:evaluation}
As described in Section~\ref{sec:attacker-model}, the goal of the attacker is to identify services on restricted networks. Based on this attacker model, we evaluate our approach on different IoT devices as well as several Content Management Systems (CMS). Furthermore, we analyzed how the approach performs in identifying parts of a composite service, in this case, WordPress plugins. 

\subsection{Identification of IoT Devices}

\begin{table*}[t]
\begin{subtable}{\columnwidth}
    \centering
    \caption{Results of Phase 1}
    \begin{tabularx}{\textwidth}{p{1.2cm}>{\raggedleft\arraybackslash}p{2cm}>{\raggedleft\arraybackslash}X}
        \toprule
        \textbf{Match} & \textbf{\# Devices} & \textbf{Percentage} \\
        \midrule
        Unique & 154 & 62\% \\
        Multiple & 61 & 24\% \\
        No & 35 & 14\%\\
        \midrule
        Total & 250 & 100\% \\
        \bottomrule
    \end{tabularx}
\end{subtable}%
\hfill
\begin{subtable}{\columnwidth}
    \centering
    \caption{Results of Phase 2}
    \begin{tabularx}{\textwidth}{p{1.3cm}>{\raggedleft\arraybackslash}p{2cm}>{\raggedleft\arraybackslash}X}
        \toprule
        \textbf{Match} & \textbf{\# Devices} & \textbf{Percentage} \\
        \midrule
        Unique & 31 & 74\% \\
        Multiple & 7 & 17\% \\
        Incorrect & 4 & 9\% \\
        \midrule
        Total & 42 & 100\% \\
        \bottomrule
    \end{tabularx}
\end{subtable}
\end{table*}

Some IoT devices run web services, for example, for configuration purposes. Furthermore, they regularly provide security-critical services and often contain critical security vulnerabilities. Some IoT devices are thus easy to attack once an attacker gains network access, which is why they are often hidden behind perimeters.

As IoT devices regularly run their software stack on specific hardware and setting up a local test lab with several IoT devices is expensive and time-consuming.  On the other hand, using productive private networks as a testbed would raise ethical concerns. Performing the JavaScript-based
identification in the wild means forcing peoples' browsers to scan their private networks to identify available
devices. This scan could lead to problems inside their network and may reveal sensitive information
about the user's local network to us without the user's knowledge or consent. 

We thus decided to use the well-known public database and search engine Shodan\footnote{https://www.shodan.io/}
to automatically create testbeds of publicly accessible IoT devices.

The service identification and evaluation process itself has to take place inside a browser. To verify the process against
a large testbed would require us to initiate every test run in a browser by hand. Instead, we automated the process by
extending the Selenium-based~\cite{seleniumhq} test suite described in Section~\ref{sec:extract_feature_vectors} to perform test runs and
return the results.
We split our evaluation process for IoT devices into two phases.

\textit{Phase 1:} The first evaluation step is intended to analyze if a feature vector generated using a crawler matches the specific device when the resulting vector is used with JavaScript. To achieve this evaluation, we crawled  embedded devices from Shodan, resulting in a data set of 250 unique feature vectors. These devices were picked from search results for different Vendors and device types. For gathering the data set, we used search queries 
from the format \texttt{"<vendor> 200 ok port:'80, 8080, 8081'"}. 
After that, we aggregate these features into feature vectors. The resulting tree consists of 250 feature vectors.

Our automated Selenium-based test suite then uses this tree to identify every device that has a feature vector in the database. We were able to recognize 62\% (154 devices) of the
250 devices. Additionally, 24\% (61 devices) were correctly classified, but their feature vectors matched multiple devices.
Multiple matches happen when the publicly accessible files of multiple devices are identical -- for example,
two device models of the same manufacturer that share the same firmware. 14\% (35 devices) of the 250 devices could not be classified; these
include 17 devices for which the crawler could identify no media files, and we thus could not create a working feature vector.

\textit{Phase 2:} The second phase evaluates if a feature vector generated from a device recognizes other devices of the same type running the same software.
To evaluate this, we queried Shodan for devices serving the same software version and verified them manually to get at least eight devices of the same device type and firmware version as a ground truth.

Based on that, we used one of those devices to build a feature vector. The remaining devices were saved to the database as testing devices.
This enables us to apply the identification process to the remaining devices using our Selenium-based test suite and report the results back to the database. 
Overall, we tested 42 devices out of which 31 were correctly classified, four vectors failed, and seven returned correct results but matched multiple devices.

\subsection{Identification of Content Management Systems}
To evaluate our techniques against Content Management Systems, we chose four major systems: WordPress, Drupal, Joomla, and TYPO3. The resulting data set contains 18 major, 219 minor, and 950 patch level versions.

The results of this evaluation are described in \autoref{tbl:cmsresults}. Using our techniques, we were able to extract distinguishable feature vectors for 100\% of the major and minor versions of WordPress, Joomla, and TYPO3.
The identification rate for minor versions of Drupal only amounts to 12/139 (9\%). This is caused by the fact that Drupal does not have patch level versions for Drupal 5 to 7. For Drupal 4 and 8, which are using patch level versions, the detection rate for minor version is 11/15 (73\%).

Usually, changes in major and minor versions of CMSs results in different feature vectors, but between patch level versions, there are no or only small changes to media files. Due to this fact, we could only distinguish 122 of 950 (13\%) precise patch level versions, because many of these patch level versions are indistinguishable. We, therefore, also measured the average cluster size of the results, i.e., the amount of patch level versions that have an identical feature vector. On average, the cluster size is 3.5, which means that \NAME{} can narrow down the set of possible patch level versions of a target down to 3 or 4.

As mentioned in Section~\ref{sec:optimization}, we optimized the identification process using a tree-based approach. This enables us to perform the identification within in best case five and worst-case 14 requests. The average request count is ten requests.
\begin{table*}[t]
\begin{center}
\caption{Results of CMS version identification. }
\label{tbl:cmsresults}
\begin{tabularx}{\textwidth}{p{2.5cm}p{1.2cm}p{2.5cm}p{1.2cm}p{5cm}>{\raggedleft\arraybackslash}X}
    \toprule
    \textbf{CMS} & 
    \multicolumn{2}{l}{\textbf{Minor versions}} & 
    \multicolumn{2}{l}{\textbf{Patch level}} &
    \textbf{Average Cluster Size} \\
    \midrule
    Joomla    & 12/12   & (100\%) & 26/102  & (25\%) & 2.17 \\
    Wordpress & 31/31   & (100\%) & 55/321  & (17\%) & 3.09 \\
    TYPO3     & 37/37   & (100\%) & 31/285   & (11\%)  & 3.61 \\
    Drupal    & 12/139  & (9\%)  & 10/242  & (4\%) & 6.05 \\
    \midrule
    Total     & 92/219 & (42\%)  & 122/950  & (13\%) & 3.5 $\varnothing$ \\
    \bottomrule
\end{tabularx}
\end{center}
\end{table*}
To measure how our techniques perform in real-world like scenarios, we performed the 
identification process on installed systems with the feature vectors learned before.
Tools for identifying CMSs and their versions on the Internet
already exist~\cite{fp-tools-plecost, fp-tools-whatweb}. Unfortunately, most of the tools
available are not maintained, were build for already deprecated CMS versions, and
are thus unreliable. Therefore, they do not provide usable ground truth, making the
evaluation of the identification process against real-world systems difficult.

We decided to create a Docker-based WordPress test lab because WordPress provides well documented official docker images. This setup contains 46 different versions available on Docker Hub\footnote{https://hub.docker.com/\_/wordpress/}. 
\NAME{} can distinguish all major and minor versions of the 46 tested WordPress versions. Additionally, seven versions (15\%) can be identified down to its patch level version.

\subsection{Identification of Vulnerable WordPress Plugins}

While searching for vulnerable targets on a network behind a security perimeter, knowledge about potentially vulnerable plugins running on a specific service is valuable for the attacker. The most widely used plugin-based system is Wordpress, which is why we focused on that.
We crawled the WPScan Vulnerability Database\footnote{\url{https://wpvulndb.com/}} for plugin versions containing vulnerabilities that are likely to be exploitable using CSRF. We considered the vulnerability classes Remote Code Execution (RCE), Cross-Site Scripting (XSS), and SQL Injection to be CSRF exploitable. 

In total, we came up with 600 vulnerable Wordpress plugins. We also added the last version before the vulnerability was introduced, all vulnerable versions, and the
first non-vulnerable version of a plugin to our corpus, resulting in 1,814 different plugin versions.
After extracting the feature vectors, \NAME{} was able to identify the existence of 598 of the 600 plugins in a particular Wordpress installation.

When identifying vulnerable plugin versions down to the minor versions, we can identify 590 of the collected 1,814 plugin versions. The relatively poor performance is due to the fact
that in many cases, the vulnerabilities were fixed in the actual plugin logic, but the security patch did not change files accessible to our techniques.
For 48 plugins, it was possible to precisely determine the versions before and after a vulnerability enabling an attacker to detect 65 vulnerabilities: 34 Cross-Site Scriptings,
12 SQL Injections, two Remote Code Executions, and 17 other vulnerabilities.

\subsection{Comparison of Feature Types}
\begin{figure*}[ht!]
    \centering
    \includegraphics[width=0.8\textwidth]{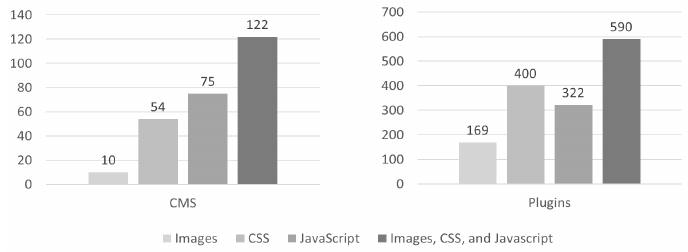}
    \caption{Comparison the effectiveness of the three different feature types.}
    \label{fig:compare-feature-types}
\end{figure*}

In Section~\ref{sec:identify}, we identified three main feature types: Images, Java\-Script, and style sheet files. We found that each
of these feature types was able to identify different services. To compare the effectiveness of each feature type, as well as the effectiveness of combinations, we measured how many CMS versions, as well as plugin versions, could
be identified correctly. \autoref{fig:compare-feature-types} shows the results.

We found that using only images as a metric, as proposed by John Bergbom \cite{forcepoint-research-report}, yields considerably worse
results than utilizing further metrics. Using images, we were still able to distinguish 169 of the 1,814 plugin versions, but different versions of
most of the Content Management System versions were no longer distinguishable. However, using either only CSS or only JavaScript files as features
improves the identification results significantly, already allowing to distinguish 54 and 75 patch level versions of CMS Systems and
400 and 322 plugin versions.

\section{Countermeasures}\label{sec:countermeasures}

In the following, we discuss several countermeasures to prevent the attacks discussed in this paper.

\subsection{Proposed Defenses for Cross-Origin Information Leaks}

Artur Janc surveyed various techniques to mitigate cross-origin information leakages \cite{artur-janc-cross-origin-infoleaks}. He references a broader range of methods, from which two seem promising to prevent the kind of attack described in our paper: \emph{Cross-Origin-Resource-Policy} and \emph{Sec-Metadata}.

\textit{Cross-Origin-Resource-Policy.}
Anne van Kesteren and John Wilander proposed to resurrect the \emph{From-Header} proposal \cite{w3c-from-origin} as \emph{Cross-Origin-Resource-Policy (CORP)} \cite{whatwg-fetch-issue-687}.
The overall idea of CORP is that a server defines which origins are allowed to \emph{process} a resource. The server does this by sending a \texttt{from-origin} header along with the requested resource. After receipt, the browser checks if
it is allowed to process the resource, and drops it otherwise.

The aim of the proposal is not to block access to the resource, but to signal that inclusion might lead to security problems. Since an attacking website is not able to modify the headers of a requested resource and the browser is expected to respect the header, CORP can effectively block the inclusion of resources. Thus, CORP also effectively prevents the identification process described in this paper.

\textit{Sec-Metadata.}
Similar to the CORP proposal, the \emph{Sec-Metadata} proposal describes a new HTTP header. In this case, the new header is sent from the client to the server. The header contains additional information about a request's origin and enables the server to decide if it wants to serve the resource or not.
Although Sec-Metadata mitigates the service identification effectively, the privacy implications are not clear. Thus, it seems not too far fetched that anti-tracking plugins may remove the header, similarly to how ``referrer-blockers'' remove the referrer header. Thus, Sec-Metadata creates a conflict between security and privacy, which might hinder adoption.

Both proposals require code changes on the server-side and the client-side. Users who regularly update their browsers will benefit from the extensions with no additional configuration efforts. Server administrators are, however, expected to reconfigure their services to use the new mitigations.
As the adoption of novel security features is often slow\footnote{The Content Security Policy (CSP) was introduced in 2012. Seven years later, 3.4\% of HTTPS sites and 0.4\% of HTTP sites of the Alexa's Top One Million websites implement the CSP \cite{csp-top-one-million}.}, both proposals describe mid or long term countermeasures.

\subsection{General Recommendations}

Although the identification process can be mitigated with the described standards and procedures, we argue that it becomes increasingly difficult to maintain security perimeters.
Furthermore, making identification more difficult may be the wrong countermeasure, as the real problem is that devices are vulnerable to CSRF attacks. Thus, developers and administrators can harden intranet services against attacks by applying the same care to internal services as they do with services facing the public internet.

While web origins as a basis for the same-origin policy are defined in RFC~6454 \cite{RFC6454}, the documentation and standardization of the same-origin policy by both standardization
bodies \cite{w3c-sop} and browser vendors \cite{mozilla-same-origin} is lacking, leading to recurring browser bugs \cite{schwenksop2018}.
Therefore, we like to encourage standardizing committees and browser vendors to find common ground in unifying the same-origin policy for a safer web.

\section{Conclusion}

A regular excuse for not securing devices in private networks is that attackers need to find those devices in the first place, which is thought to be difficult. We show that this is not true. Our \NAME{} tool is a JavaScript-based scanner that can reliably identify web service versions under the assumption that a victim opens a malicious web site from within the private network. The evaluation shows that \NAME{} can identify IoT devices running a web service, the four most widely used Content Management Systems and even vulnerable WordPress plugins with high accuracy.

\NAME{} uses techniques that leak information of cross-origin requests like image dimensions, the existence of certain CSS styles or JavaScript function names. The Same-Origin Policy implementations of Mozilla Firefox, Google Chrome, and Apple Safari allow these techniques, and they cannot be fixed without impeding benign functionality of the web.

Although countermeasures exist, for example, CORP and Sec-Metadata, not only browsers need to support them, but also server-side configuration is needed for every web service. Thus, unfortunately, no short term solutions exist, and securing the endpoints behind perimeters remains the most effective way to thwart attacks. In this case, attackers can still identify type and version of web services, but this information is useless if the service does not contain any known vulnerabilities.

\begin{acks}
The authors would like to thank Sebastian Lekis, Arthur Junk, Ben Stock and Martin Grothe for their valuable feedback and insightful discussions.
Christian Dresen was supported by the research training group ``Human Centered System Security’‘, sponsored by the state of North Rhine-Westfalia.
Fabian Ising was supported by the research project ``MITSicherheit.NRW’' funded by the European Regional Development Fund North Rhine-Westphalia (EFRE.NRW).
\end{acks}

\bibliographystyle{ACM-Reference-Format}
\interlinepenalty=10000
\balance
\bibliography{bib/main,bib/rfc}

\end{document}